\PassOptionsToPackage{unicode=true}{hyperref} 
\PassOptionsToPackage{hyphens}{url}
\documentclass[astrosymb, twocolumn, tighten]{aastex62}
\usepackage{lmodern}
\usepackage{amssymb,amsmath}
\usepackage{ifxetex,ifluatex}
\usepackage{fixltx2e} 
\ifnum 0\ifxetex 1\fi\ifluatex 1\fi=0 
  \usepackage[T1]{fontenc}
  \usepackage[utf8]{inputenc}
  \usepackage{textcomp} 
\else 
  \usepackage{unicode-math}
  \defaultfontfeatures{Ligatures=TeX,Scale=MatchLowercase}
\fi
\IfFileExists{upquote.sty}{\usepackage{upquote}}{}
\IfFileExists{microtype.sty}{%
\usepackage[]{microtype}
\UseMicrotypeSet[protrusion]{basicmath} 
}{}
\IfFileExists{parskip.sty}{%
\usepackage{parskip}
}{
\setlength{\parindent}{0pt}
\setlength{\parskip}{6pt plus 2pt minus 1pt}
}
\usepackage{hyperref}
\hypersetup{
            pdfborder={0 0 0},
            breaklinks=true}
\usepackage{listings}

\usepackage{graphicx,grffile}
\makeatletter
\def\maxwidth{\ifdim\Gin@nat@width>\linewidth\linewidth\else\Gin@nat@width\fi}
\def\maxheight{\ifdim\Gin@nat@height>\textheight\textheight\else\Gin@nat@height\fi}
\makeatother
\setkeys{Gin}{width=\maxwidth,height=\maxheight,keepaspectratio}
\ifx\paragraph\undefined\else
\let\oldparagraph\paragraph
\renewcommand{\paragraph}[1]{\oldparagraph{#1}\mbox{}}
\fi
\ifx\subparagraph\undefined\else
\let\oldsubparagraph\subparagraph
\renewcommand{\subparagraph}[1]{\oldsubparagraph{#1}\mbox{}}
\fi

\makeatletter
\def\fps@figure{htbp}
\makeatother

\usepackage[capitalize]{cleveref}
\usepackage{CJKutf8}
\newcommand{\chinesename}{{\begin{CJK}{UTF8}{gbsn}(王加冕)\end{CJK}}}
\usepackage[]{natbib}
\date{November 15, 2018}

\begin{document}

\title{Explaining Deviations from the Scaling Relationship of the Large Frequency Separation}
\correspondingauthor{Joel Ong}
\email{joel.ong@yale.edu}
\author[0000-0001-7664-648X]{Joel Ong J. M. \chinesename}
\affiliation{Department of Astronomy, Yale University, 52 Hillhouse Ave., New Haven, CT 06511, USA}
 \author[0000-0002-6163-3472]{Sarbani Basu}
\affiliation{Department of Astronomy, Yale University, 52 Hillhouse Ave., New Haven, CT 06511, USA}
\received{September 11, 2018}
\accepted{November 15, 2018}
\submitjournal{\apj}

\begin{abstract}
Asteroseismic large frequency separations possess great diagnostic value. However, their expressions as scaling relations are predicated on homology arguments which may not hold in general, resulting in mass- and temperature-dependent deviations. The first-order asymptotic expressions, which should in principle account for this structural evolution, also deviate more from fitted frequency-separation estimates than do the simple scaling relations, and exhibit qualitatively different behavior. We present a modified asymptotic estimator, and show that these discrepancies can be accounted for by the evolution of the acoustic turning points of the asteroseismic mode cavity, which is typically neglected in first-order asymptotic analysis. This permits us to use a single expression to accurately estimate the large frequency separations of main-sequence, ascending red giant branch, and red clump stellar models, except at transition points between two asymptotic regimes during the subgiant phase of evolution, where the WKB approach fails. The existence of such transition points provides theoretical justification for separately calibrated scaling relations for stars in different evolutionary stages.

\keywords{methods: analytical, methods: numerical, stars: oscillations}
\end{abstract}

\hypertarget{introduction}{%
\section{Introduction}\label{introduction}}

Solar-like oscillations --- oscillations of the star excited
stochastically in convective envelopes --- result in photometric and
velocity variations which have frequency-domain power spectra that
exhibit a comb-like structure. Such a comb is composed of peaks at the
resonant frequencies of global asteroseismic p-modes, which are (to a
good approximation) evenly spaced in the frequency domain with a
characteristic spacing \(\Delta\nu\), which is called the large
frequency separation. Phenomenologically, there exist minor variations
in the mode frequency spacing (owing to e.g.~acoustic glitches), and in
practice an average value of \(\Delta\nu\) is found by least-squares
fitting of mode frequencies against their associated radial quantum
numbers, with appropriate weights to account for variations of noise and
mode amplitudes.

The asymptotic expression for these p-mode frequencies,
\begin{equation}2 \nu_{nl} T_0 = n + {l \over 2} + \alpha, \label{eq:freqs}\end{equation}
where \(T_0 = 2\int_0^R {\mathrm d r \over c_s}\) is the sound-travel
time, yields an expression for the large frequency separation of radial
modes as \begin{equation}
    \Delta\nu_{n, 0} = \nu_{n+1,0} - \nu_{n, 0} \sim {1 \over 2 T_0}.\label{eq:soundtraveltime}
\end{equation} Under assumption of hydrostatic equilibrium, the
sound-travel time and gravitational dynamical time
\(T_\text{ff} \sim 1/\sqrt{G \rho}\) should be of roughly the same order
of magnitude, giving rise to the scaling relation
\citep{ulrich_determination_1986, christensen-dalsgaard_hertzsprung_1988}
\begin{equation}
    {\Delta\nu \over \Delta\nu_\sun} \sim \sqrt{\overline{\rho} \over \overline{\rho}_\sun},
\end{equation} with \(\overline{\rho}\) as the mean density of the star,
for stars that are essentially homologous to the Sun. This
solar-calibrated scaling relation has been historically established to
hold relatively well even for stellar models that exhibit considerable
structural differences from the Sun. For this reason, the large
separation has proven to be a valuable diagnostic of global properties
of stars exhibiting such solar-like oscillations
\citep{chaplin_asteroseismic_2014, pinsonneault_apokasc_2014, serenelli_first_2017}.

Owing to this first-order dependence on the mean density, observational
measurements of \(\Delta\nu\) have been widely applied to determine
asteroseismic estimates of stellar masses and radii, either by direct
inversion of this scaling relation
\citep[e.g.][]{chaplin_asteroseismic_2010}, or through a so-called
grid-based approach, where an optimal solution is determined from the
large separations of a grid of models
\citep[e.g.][]{basu_sounding_2011, chaplin_asteroseismic_2014, pinsonneault_apokasc_2014}.
For the latter case, it is generally desirable to be able to accurately
estimate \(\Delta\nu\) from a stellar model without solving for
individual mode frequencies, which is quite computationally expensive.
The scaling relation is typically used for this purpose.

\begin{figure}
\centering
\includegraphics{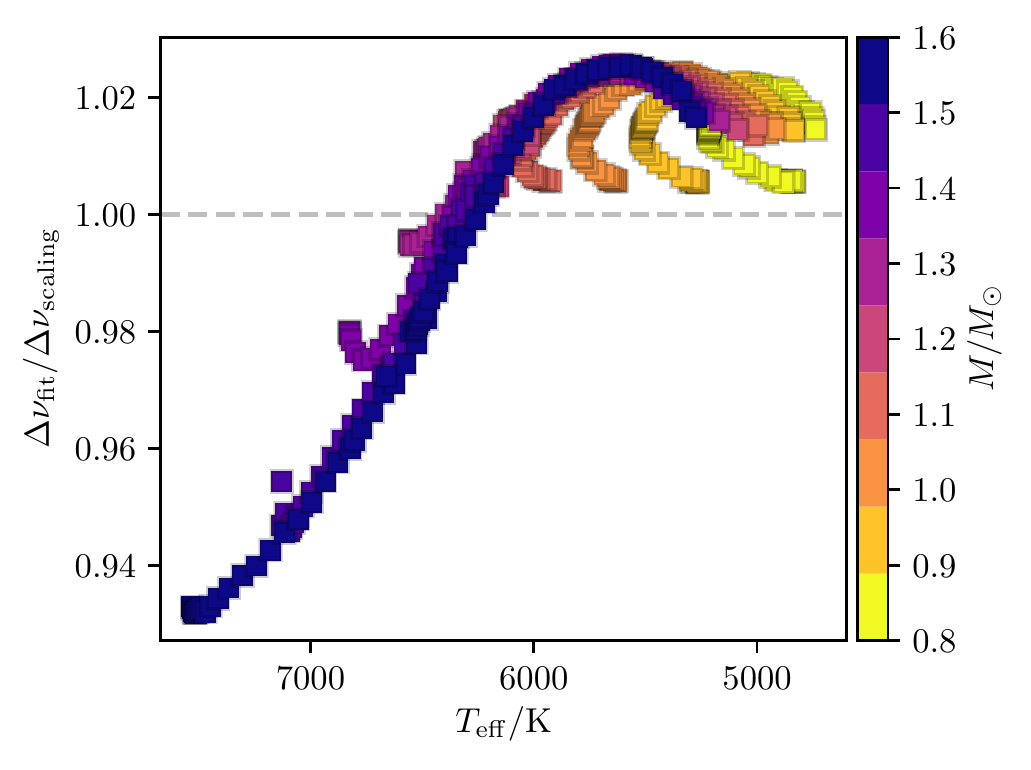}
\caption{Scaling deviations: ratio of fitted \(\Delta\nu\) versus
scaling relation estimator for solar-metallicity main-sequence/subgiant
\texttt{MESA} stellar models with solar-calibrated parameters over a
range of stellar masses.\label{fig:scaling}}
\end{figure}

However, there remain discrepancies (of order a few percent) between the
scaling relation predictions compared to fitted values of \(\Delta\nu\)
as computed from detailed frequencies returned from stellar models
(i.e.~even in the absence of observational errors and unconstrained
physics), which prevent its use in such a capacity in the regime of
high-precision asteroseismology. \citet{white_calculating_2011} find
that these scaling deviations appear to exhibit some dependence on the
effective temperature (and by proxy on stellar evolution --- see
\cref{fig:scaling}), with an apparent second-parameter dependence on
stellar metallicity.

Extant approaches to dealing with these deviations vary, but are
uniformly empirical in nature. For instance,
\citet{white_calculating_2011} fit a parabola to the empirical
morphology of the scaling deviation curve, while
\citet{guggenberger_significantly_2016} propose the use of a
damped-sinusoid function \citep[with additional modifications from
symbolic regression presented in][]{guggenberger_mitigating_2017}, and
\citet{kallinger_nonlinear_2018} propose modifying the first-order
expression \cref{eq:freqs} to include additional free parameters,
accounting for acoustic glitches and second-order curvature effects,
resulting in additional terms in the scaling relation that have to be
calibrated empirically. \citet{sharma_stellar_2016} instead seek
recourse to explicitly calibrating the scaling deviations against a
reference grid of stellar models spanning the desired parameter space.
While undoubtedly practical, such approaches lack theoretical insight.
Moreover, attempts at empirical fits (without recourse to a model grid)
have neglected scaling deviations on the main sequence, despite these
stars being in principle the most similar to the Sun (and hence
ostensibly the easiest to model).

\begin{figure}
\centering
\includegraphics{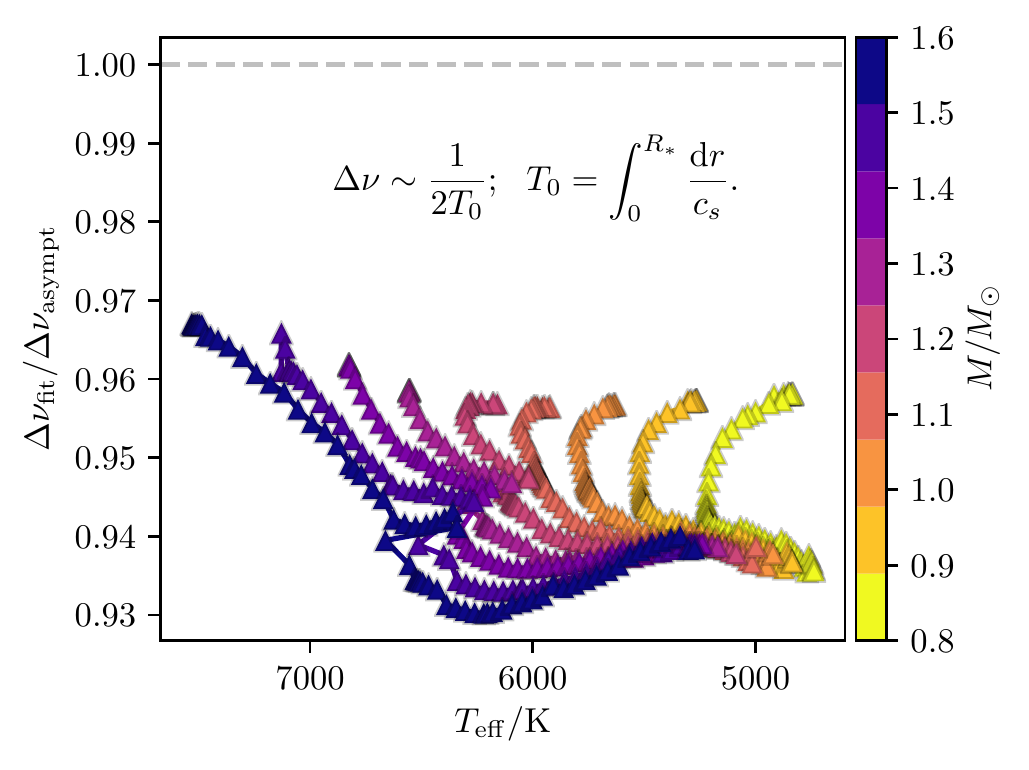}
\caption{Ratio of fitted \(\Delta\nu\) versus the usual asymptotic
estimator (i.e.~sound-travel time) for the same set of stellar models as
\cref{fig:scaling}.\label{fig:asymptotic}}
\end{figure}

While the sound-travel time may also be used to estimate \(\Delta\nu\)
without requiring individual mode frequencies, there exists some
contention as to the correct choice of the upper limit \(R\) used to
compute it. Strictly speaking, the limit of integration should be the
outer boundary of the eigenvalue problem with respect to which the
eigenfunctions are computed. However, this boundary is difficult to
define. While the photospheric radius is used by default in most
evolutionary codes, it returns sound-travel times that are typically
lower (and therefore frequency separations higher) than would be
consistent with both observational values and values fitted against
detailed model frequencies (as in \cref{fig:asymptotic}).
\citet{hekker_tests_2013} note that this can be remedied by simply
extending the domain of integration outwards, but the question of
ambiguity as to the correct outer boundary remains.

A critical underlying assumption made in all of these analyses is that
the phenomenological and asymptotic values of \(\Delta\nu\) should be
similar, which is only true if the phase function \(\alpha\) in
\cref{eq:freqs} does not exhibit significant secular variation between
modes. This would be the case with e.g.~the contribution to \(\alpha\)
from an acoustic glitch, where the rapid but oscillatory variations with
frequency are cancelled out when a global value of \(\Delta\nu\) is
fitted for. However, in this formulation, the value of \(\alpha\) as a
function of frequency depends on the structure of the entire stellar
model, and this assumption may not necessarily hold good. On the other
hand, should such secular phase variations exist, their dependence on
stellar parameters like the mass and radius is not \emph{a priori}
obvious.

We derive instead an asymptotic estimator for the large frequency
separation which captures most of the variation in these phase functions
(and so scaling deviations) with a single expression, and thereby
returns estimates of \(\Delta\nu\) that are considerably closer to the
fitted value than the traditional asymptotic estimator, without any
ambiguity as to the outer turning point of the relevant integral. We
demonstrate numerically that the validity of this expression is
independent of stellar mass, evolutionary stage (with one major
exception), and atmospheric model.

\hypertarget{the-wkb-approximation}{%
\section{The WKB approximation}\label{the-wkb-approximation}}

\hypertarget{review-of-existing-work}{%
\subsection{Review of existing work}\label{review-of-existing-work}}

In the first-order asymptotic theory of p-modes, the large frequency
separation \(\Delta\nu\) emerges from the phase integral quantization
condition
\begin{equation}\int_{r_1}^{r_2}{k_r(r, \omega)}{\mathrm d r} = \pi \left(n + \kappa\right)\end{equation}
for integer \(n\), where \(k_r\) is the WKB wavenumber, \(r_1\) and
\(r_2\) are the turning points of the integral (usually where \(k_r\)
vanishes), and \(\kappa\) is a function of frequency. Here, \(\kappa\)
depends only on the functional behavior of \(k_r^2\) at the turning
points. In particular it can be shown that \(\kappa = -{1 \over 2}\) is
constant if \(k_r^2 \propto r-r_t\) at both turning points
\citep{gough_elementary_2007, aertsbook} . To a first approximation,
this expression yields the Duvall law,
\begin{equation}\int_{r_t}^{R} {\mathrm d r \over c_s} \sqrt{1 - {S_l^2 \over \omega^2}} = \pi \left({n  + \kappa \over \omega}\right),\label{eq:duvall}\end{equation}
where \(c_s\) is the sound speed, \(S_l^2 = {l(l+1) \over c_s^2 r^2}\)
is the Lamb frequency, and \(\omega = 2\pi\nu\) is the angular
frequency. Again the inner turning point of this integral is defined by
where the integrand vanishes (or \(r=0\) for radial modes), and the
outer turning point \(R\) is the same as that used to evaluate the
sound-travel time.

Naturally, higher-order approximations are possible. For instance,
\citet{tassoul_second-order_1994} derives \cref{eq:freqs} from a
second-order expansion of the asteroseismic equations of motion, using
linear combinations of spherical Bessel functions and their derivatives
as ansatzen for the eigenfunctions, with a similar phase integral
quantization condition emerging for the arguments of these functions.
This was in turn taken to fourth order by
\citet{roxburgh_asymptotic_1994} with a similar approach.

On the other hand, the accuracy of the WKB expression can also be
improved with more detailed asymptotic analysis (i.e.~not setting terms
to zero prematurely before performing the WKB analysis). For instance, a
more accurate description is afforded by
\citet{deubner_helioseismology:_1984}:
\begin{equation}\int_{r_1}^{r_2} {\mathrm{d} r \over c_s} \sqrt{1 - {\omega_\text{ac}^2 \over \omega^2} -{S_l^2 \over \omega^2}\left(1 - {N^2 \over \omega^2}\right)} = \pi \left({n + \kappa' \over \omega}\right),\label{eq:jwkb}\end{equation}
which emerges from just such a more detailed analysis \citep[with still
more detail provided in][]{gough_linear_1993}. This integrand in this
expression reduces asymptotically to that in \cref{eq:duvall} far into
the interior of the convective envelope of solar-like main-sequence
stars, where \(\omega_\text{ac} \ll \omega\) for most p-modes of
interest. Here \(N^2\) is the Brunt-Väisälä or buoyancy frequency, and
\(\kappa'\) is a another function of frequency (different in general
from \(\kappa\) in \cref{eq:duvall}). However, \(\kappa'\) also only
depends on the behavior of the integrand in \cref{eq:jwkb} at the
turning points.

The formulation of \cref{eq:jwkb} differs from the standard form of the
Duvall law both in terms of its explicit dependence on an acoustic
cutoff frequency
\begin{equation}\omega_\text{ac}^2 = {c_s^2 \over 4 H^2}\left(1 - 2{\mathrm d H \over \mathrm d r}\right),\label{eq:ac}\end{equation}
involving \(H=-\left(\mathrm d \ln \rho \over \mathrm d r\right)^{-1}\)
(the density scale height) in the integrand, and in terms of the
locations of the turning points (as the integrands vanish at different
places). In particular, the outer turning point is essentially
determined by the acoustic cutoff frequency, rather than being defined
by the boundary conditions of the eigenvalue problem. The precise form
of the acoustic cutoff frequency, and other quantities involved in the
oscillation equations, will depend on the choice of dynamical variable
used to perform the WKB analysis. Nonetheless, the resulting eigenvalue
equation can typically be put into the form of \cref{eq:jwkb}, although
different expressions for \(\omega_\text{ac}\), \(S_l^2\) and \(N^2\)
may have to be used in place of those described here. An illustrative
example can be found in \citet{gough_linear_1993}.

With respect to the above formalism,
\citet{christensen-dalsgaard_phase_1992} instead define a phase function
\(\alpha\) so that \begin{equation}
    \int_{r_t}^{R} {\mathrm d r \over c_s} \sqrt{1 - {S_l^2 \over \omega^2}} = \pi \left({n  + \alpha\over \omega}\right),
 \end{equation} where \begin{equation}
    \alpha(\omega) = {\omega \over \pi} \left(T_1 - T_2\right) + \kappa',
\end{equation} with \begin{equation}
    \begin{aligned}
T_1 &= \int_{r_t}^{R} {\mathrm d r \over c_s}\sqrt{1 - {S_l^2 \over \omega^2}},\\
T_2 &= \int_{r_1}^{r_2}{\mathrm d r \over c_s}  \sqrt{1 - {\omega_\text{ac}^2 \over \omega^2} - {S_l^2 \over \omega^2}\left(1 - {N^2 \over \omega^2}\right)}
\end{aligned}
\end{equation} being integrals with dimensions of time appearing on the
LHS of \cref{eq:duvall,eq:jwkb}. This permits the use of \cref{eq:freqs}
with \(\alpha\) in place of \(\kappa\). Whereas the phase functions
\(\kappa\) and \(\kappa'\) defined previously depend only locally on the
asymptotic properties of the WKB integrands in the neighborhoods of
their classical turning points, \(\alpha\) as defined in this manner
depends also on global properties of the entire acoustic mode cavity.

\hypertarget{a-new-expression}{%
\subsection{A new expression}\label{a-new-expression}}

We take the first-order expression \cref{eq:jwkb} at face value, and
assume that the quantities \(N^2\) and \(S_l^2\) have no parametric
dependence on the frequency, and further that all quantities on the RHS
are continuous functions of the frequency \(\nu\). We proceed to expand
finite differences as Taylor series. In particular, we consider finite
differences of the terms in \cref{eq:jwkb}, multiplied throughout by
\(\omega/\pi = 2\nu\) and evaluated at frequencies of adjacent \(n\) for
the same \(l\):
\begin{equation}\begin{aligned}2\nu_{n+1,l}T_2(\nu_{n+1,l}) &- 2\nu_{n-1,l}T_2(\nu_{n-1,l}) \\&\sim 2 +  \kappa'\left(\nu_{n+1, l}\right) - \kappa'\left(\nu_{n-1, l}\right) \\ \implies 4{\mathrm d \nu T_2 \over \mathrm d \nu} \Delta\nu &\sim 2 + 2{\mathrm d \kappa' \over \mathrm d \nu}\Delta\nu,\end{aligned}\end{equation}
with the quantities on the last line being evaluated at the frequency
\(\nu = {1 \over 2}(\nu_{n+1,l} +\nu_{n-1,l})\) so that the error term
is \(\mathcal{O}(\Delta\nu^3)\), and where \(T_2\) is the LHS of
\cref{eq:jwkb}. This yields the expression
\begin{equation}\Delta\nu \sim \left(2 {\mathrm d \nu T_2 \over \mathrm d \nu} - {\mathrm d \kappa' \over \mathrm d \nu}\right)^{-1}.\label{eq:intermediate1}\end{equation}

For the first term, we observe that
\begin{equation}\begin{aligned}&{\mathrm d \over \mathrm d \nu} \left(\nu \int_{r_1}^{r_2}\sqrt{f(r, \nu)}~{\mathrm d r \over c_s}\right) = {\mathrm d \over \mathrm d \omega} \left(\omega \int_{r_1}^{r_2}\sqrt{f(r, \omega)}~{\mathrm d r \over c_s}\right) \\ &= \int_{r_1}^{r_2}{f(r, \omega) + {1 \over 2}\omega{\partial \over \partial \omega}f(r, \omega) \over \sqrt{f(r, \omega)}}~{\mathrm d r \over c_s} \\ &+ \omega\left({\sqrt{f(r_2, \omega)} \over c_s}{\mathrm d r_2 \over \mathrm d \omega} - {\sqrt{f(r_1, \omega)} \over c_s}{\mathrm d r_1 \over \mathrm d \omega}\right).\end{aligned}\end{equation}
Since the turning points of the integral are either fixed (for
\(r_1=0\)) or defined to be where the relevant integrand vanishes, the
boundary contributions to the above derivative also vanish, leaving us
with
\begin{equation}2 \int_{r_1}^{r_2}{\mathrm d r \over c_s}{1 - {S_l^2 N^2 \over \omega^4}\over \sqrt{1 - {\omega_\text{ac}^2 \over \omega^2}- {S_l^2 \over \omega^2}\left(1 - {N^2 \over \omega^2}\right)}} - {\mathrm d \kappa' \over \mathrm d \nu}\end{equation}
inside the parentheses in \cref{eq:intermediate1}. For radial modes in
particular, we can neglect \(S_l^2\), leaving us with
\begin{equation}\Delta\nu \sim \left(2 \int_{r_1}^{r_2}{\mathrm d r \over c_s}{1 \over \sqrt{1 - {\omega_\text{ac}^2 \over \omega^2}}} - {\mathrm d \kappa' \over \mathrm d \nu}\right)^{-1}.\end{equation}

As noted previously, \(\kappa' = -{1 \over 2}\) in the ideal WKB
scenario of the mode cavity exhibiting asymptotically linear behavior
near the classical turning points. On the other hand, if the radicand
tends to a constant value near the inner turning point (as would be the
case for Sun-like stars, where \(\omega_{ac} \ll \omega\) far into the
interior), then the behavior of the eigenfunctions near the inner
turning point are described by the spherical Bessel function \(j_0\),
and \(\kappa' = -{1 \over 4}\) \citep{gough_linear_1993}. In either
case, assuming that it changes very slowly with respect to frequency,
its derivative is much smaller than the integral expression. We thus
propose the use of a modified integral estimator for the large frequency
separation:
\begin{equation}\Delta\nu \sim \left(2 \int_{r_1}^{r_2}{\mathrm d r \over c_s}{1 \over \sqrt{1 - {\omega_\text{ac}^2 \over \omega^2}}}\right)^{-1}.\label{eq:newexpr}\end{equation}
This expression contains an explicit dependence on the frequency.
Observationally, however, we are limited to the study of only modes with
frequencies near \(\nu_\text{max}\), the frequency of maximum acoustic
power. Moving forward, we will evaluate this quantity at
\(\nu_\text{max}\).

We note that the denominator of the integrand in \cref{eq:newexpr}
vanishes at the outer boundary of the domain of integration, so the
integrand becomes singular. Depending on the acoustic cutoff frequency
of the stellar model, this may also be the case at the inner boundary.
In the ideal WKB regime, this results in the presence of integrable
singularities at the endpoints. We would therefore expect that the
evolution of these turning points plays a significant role in how the
value of this estimator changes over the course of stellar evolution.
Our expression does not contain a prescription for the actual turning
points of the asymptotic integral, only that they depend on the
formulation of the acoustic cutoff frequency that accompanies the degree
of approximation being used.

Moreover, within the domain of integration, we note that the integrand
is strictly larger than the \(1/c_s\) that appears as the integrand of
the sound-travel time. Since the integrand is also strictly positive, we
conclude that the integral is strictly larger than the sound-travel
time. Given that the sound-travel time generally overestimates
\(\Delta\nu\) (i.e.~yields too small a value of the integral, as in
\cref{fig:asymptotic}), this is suggestive as at least a zeroth-order
mark of improvement.

\hypertarget{benchmarking-against-models}{%
\section{Benchmarking against
models}\label{benchmarking-against-models}}

\begin{figure*}[p]
\centering
\includegraphics{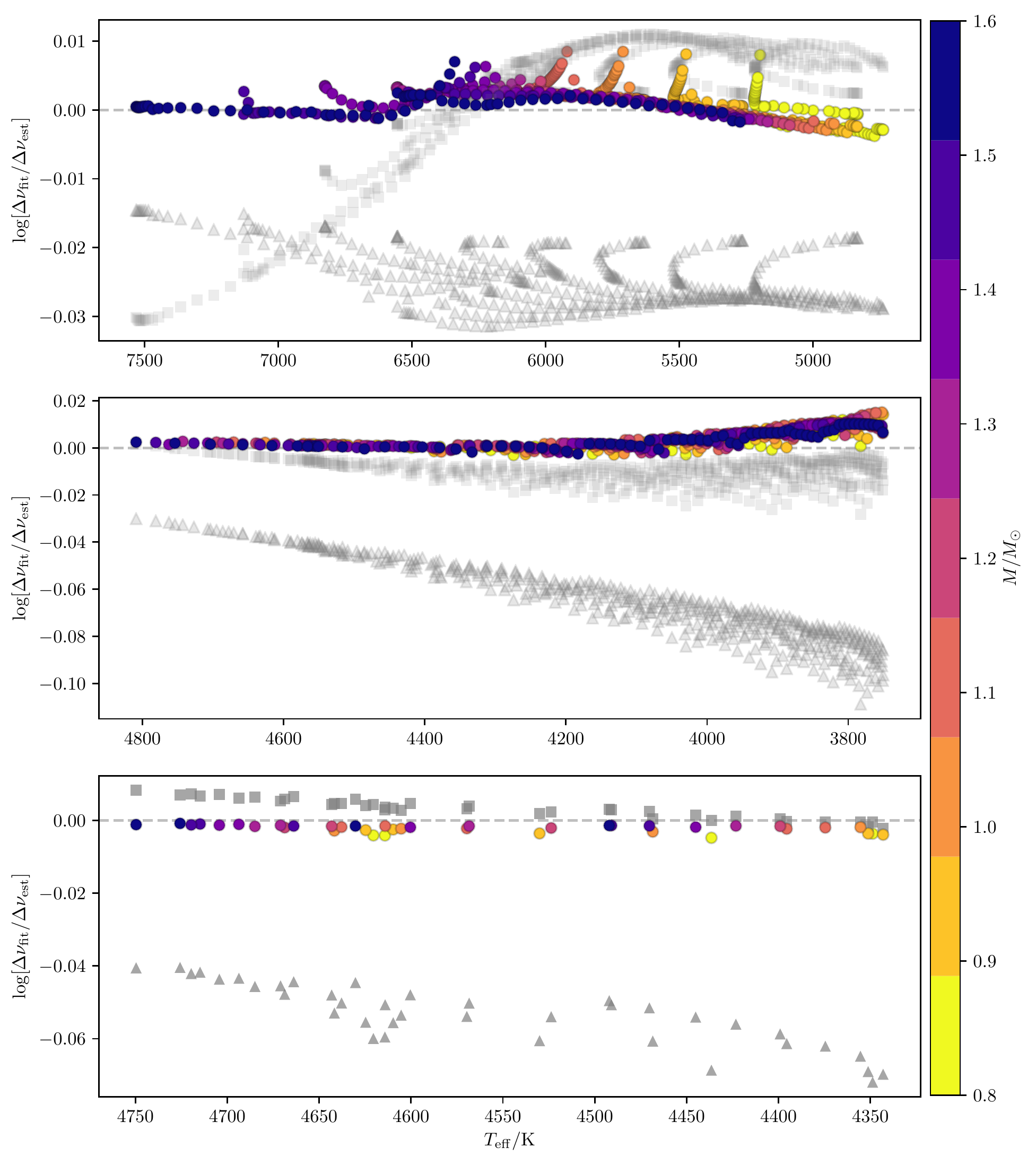}
\caption{Log ratio of fitted \(\Delta\nu\) versus various estimators, for solar-metallicity \texttt{MESA} stellar models with solar-calibrated parameters over a range of stellar masses and ages, with our modified asymptotic estimator in colored points. To guide the eye, we plot with faint markers the quantities in \cref{fig:scaling,fig:asymptotic} (with corresponding marker shapes).  We break these plots up by evolutionary stage. \textbf{Top:} main-sequence and subgiant stars; \textbf{Middle:} Ascending RGB; \textbf{Bottom:} Horizontal branch/red clump models. The agreement is considerably better for our estimator than the usual methods.\label{fig:quadrature}}
\end{figure*}

Our modified estimator for \(\Delta\nu\), \cref{eq:newexpr}, was derived
under assumptions that may not be strictly valid in all cases. To test
how well it actually reproduces the large separation of a model, we
compare it against \(\Delta\nu\) as calculated using the frequencies of
a model.

To do this, we compute both \cref{eq:newexpr} as well as an estimate of
\(\Delta\nu\) from a linear fit to individual mode frequencies, for a
variety of stellar models constructed with the stellar evolution code
\texttt{MESA}, version 10398 \citep{paxton_modules_2015}. We ran
evolutionary tracks for stellar masses from 0.8 to 1.6 solar masses in
increments of \(0.1 M_\Sun\), using initial helium abundances \(Y_0\)
and mixing-length parameters \(\alpha_\text{MLT}\) obtained by
constructing solar-calibrated models at solar metallicity \citep[using
values from][]{grevesse_standard_1998}, but with no additional
constraints. Frequencies were computed with version 5.1 of the
\texttt{GYRE} oscillation code \citep{townsend_gyre:_2013}. The
evolution included heavy element diffusion and gravitational settling as
well as convective overshoot. We compute frequencies for all radial
modes up until the maximum acoustic cutoff frequency (i.e.~all trapped
p-modes)

As noted above, our proposed asymptotic estimator has an explicit
frequency dependence, and we evaluate it at the frequency of maximum
acoustic power, \(\nu_\text{max}\). While the theoretical underpinnings
of \(\nu_\text{max}\) are not well-understood, it carries diagnostic
information on the excitation and damping of stellar modes, and hence
must depend on the physical conditions in the near-surface layers where
the modes are excited. Close to the surface (and outer acoustic turning
point), the behavior of the waves is strongly influenced by the acoustic
cut-off frequency. \citet{brown_detection_1991} argue that
\(\nu_\text{max}\propto\nu_\text{ac}\), since both frequencies are
determined by conditions in the near-surface layers;
\citet{kjeldsen_amplitudes_1995} use this to derive a scaling relation
between \(\nu_\text{max}\) and near-surface properties.
\citet{belkacem_underlying_2011} show that while \(\nu_\text{max}\) does
indeed depend on \(\nu_\text{ac}\), there are also additional
dependences on other quantities, such as the turbulent Mach number and
the mixing-length parameter. However, to simplify matters, in this work
we evaluate \cref{eq:newexpr} at the value of \(\nu_\text{max}\) given
by the the scaling relation of \citet{kjeldsen_amplitudes_1995}.

The expression \cref{eq:ac} for the acoustic cutoff frequency exhibits
rapid variations, both owing to limitations in the MESA models and
possibly density discontinuities, which may affect the accuracy of the
WKB approximation (in that \(k_r\) may not vary slowly). In particular,
large rapid variations near the turning points will strongly influence
the computed value of the integrable singularities there. To avoid this,
we use an isothermal homogeneous plane-parallel ideal-gas approximation,
as detailed in \citet{aertsbook}. Practically, this means that we use
only the leading order term in \cref{eq:ac} when computing the acoustic
cutoff frequency. In turn, the value of the integral \cref{eq:newexpr}
was evaluated using an adaptive Gauss-Kronrod quadrature scheme.

\hypertarget{results}{%
\subsection{Results}\label{results}}

To compare the results obtained from the integral and from frequencies,
we compute the logarithms of the ratios between the value of
\(\Delta\nu_\text{fit}\) (obtained by a least-squares fit of
\(\nu_{n, l=0}\) against \(n\)), versus the values of our integral
estimator in \cref{eq:newexpr}, which we plot as circles in subsequent
figures. We also do this with the log ratios of \(\Delta\nu_\text{fit}\)
versus the values of the sound-travel time estimator \(1/2T_0\) (which
we show as upright triangles), as well as versus the values predicted by
the scaling relation (squares). As a more quantitative (but still
heuristic) comparison, we compute the root-mean-square (RMS) deviation
of these log-ratios from zero. An RMS deviation of zero means that the
estimator exactly coincides with \(\Delta\nu_\text{fit}\) for all
models.

The top panel of \cref{fig:quadrature} shows the values of the log
ratios described above for stellar models along the main sequence and on
the subgiant branch. It is visually evident that the agreement between
the estimator in \cref{eq:newexpr} and the fitted values of
\(\Delta\nu\) is considerably better than both the scaling relation and
the sound-travel time. We find that the scaling relation results in a
RMS log-deviation of about 0.010 dex, compared to 0.025 dex for the
sound-travel time and 0.002 dex for our integral estimator.

\begin{figure}[hbt]
\centering
\includegraphics{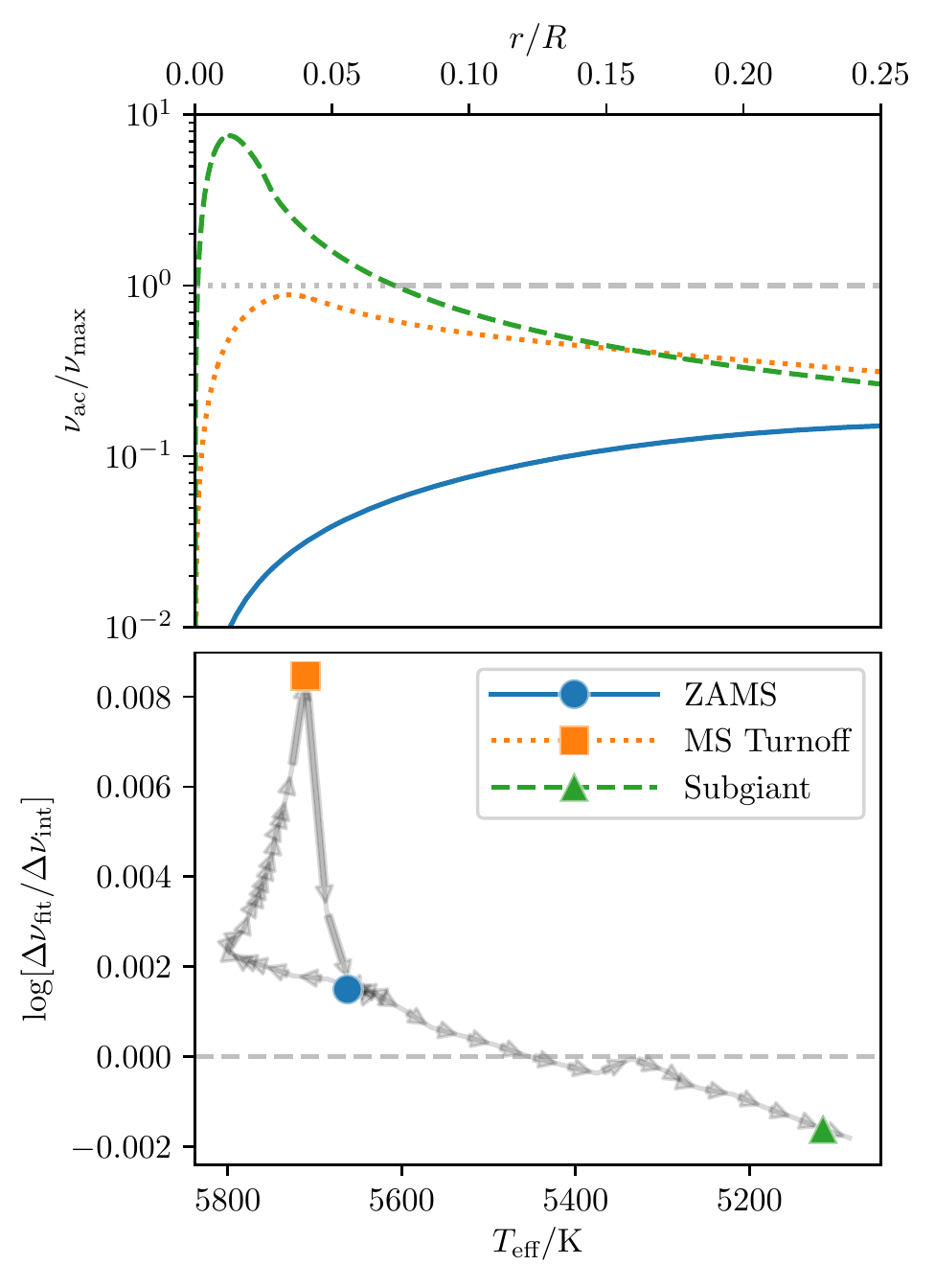}
\caption{\textbf{Top:} Plots of the acoustic cutoff frequency in units of $\nu_\text{max}$ for stellar models at various evolutionary stages along the $1 M_\sun$ track, showing the two extreme asymptotic regimes of the inner turning points for main-sequence and evolved stars, and illustrating the failure of the WKB approximation during the transition between them at main-sequence turnoff. \textbf{Bottom:} Log-ratio of frequency separations (as in top panel of \cref{fig:quadrature}) for $1 M_\sun$ track, with direction of evolution indicated with arrows. Stellar models corresponding to curves shown in the top panel are marked out with points of the same color, on lines with corresponding linestyles, in the legend of the bottom panel.\label{fig:cavity}}
\end{figure}

Whereas the relative deviations from the scaling relation appear to
exhibit curlicues for main-sequence stars of different masses (as in
\cref{fig:scaling}), we note that the most obvious deviations between
our asymptotic estimator and the fitted value of \(\Delta\nu\) occur as
a sharp spike near main-sequence turnoff during the onset of shell
burning (at core hydrogen exhaustion/subgiant hook for more massive
stars). We examine this more closely with \cref{fig:cavity}.

We show in the top panel of \cref{fig:cavity} the acoustic cutoff
frequency, in units of \(\nu_\text{max}\), of main-sequence, MS turnoff,
and subgiant stellar models with \(M=1.0M_\sun\). Following our above
discussion of various values of \(\kappa\) in the WKB regime, we see
that for the main-sequence star (blue solid curve), the inner turning
point is at \(r=0\), where \(\nu_\mathrm{ac}/\nu_\text{max} \ll 1\), and
the WKB wavefunction goes as \(j_0(\omega t)\), where
\(t \sim \int_0^r \sqrt{1 - \left(\omega_\text{ac}/\omega\right)^2} {\mathrm{d}r \over c_s}\),
with the same integrand as in \cref{eq:jwkb}. On the other hand, for the
subgiant star (green dashed curve), the WKB wavefunction goes as
\(\mathrm{Ai}(- |\omega t|^{3 \over 2})\) near a first-order classical
turning point at \(r = r_1\), where
\(\nu_\mathrm{ac}(r_1) \sim \nu_\text{max}\) and
\(t \sim \int_{r_1}^r \sqrt{1 - \left(\omega_\text{ac}/\omega\right)^2} {\mathrm{d}r \over c_s}\),
and \(\mathrm{Ai}\) is the Airy function.

These represent two distinct asymptotic regimes in which the assumptions
underlying the WKB approximation hold well. However, this is not
necessarily true during the transition between these two regimes. In
particular, for some stellar models near main-sequence turnoff (yellow
dotted curve), we have a maximum of the acoustic cutoff frequency with a
value very close to \(\nu_\text{max}\). This violates one of the
underlying assumptions of the WKB approximation (viz. that \(k_r\)
varies slowly except near classical turning points, which requires in
this case that \(\nu \gg \nu_\mathrm{ac}\)), and neither of the above
asymptotic descriptions is a good approximation for the actual mode
eigenfunctions.

We have earlier noted that the radicand in \cref{eq:newexpr} vanishes
where \(\nu_\mathrm{ac} = \nu_\text{max}\). For first-order turning
points, this results in integrable singularities at the endpoints of
integration. However, when there is a maximum point where
\(\nu_\mathrm{ac}/\nu_\text{max} \sim 1\) (as would be the case with a
second-order turning point), the integrand either changes very rapidly
(becoming very large) in a region smaller than the wavelength of the
eigenfunction (\(\max \nu_\mathrm{ac}/\nu_\text{max} \lessgtr 1\)), or a
nonintegrable singularity is introduced into the domain of integration
(\(\max \nu_\mathrm{ac}/\nu_\text{max} = 1\)), and the convergence of
the numerical integration is poor. In these cases, a numerical
computation of \cref{eq:newexpr} will be very different from the fitted
value of \(\Delta\nu\): the yellow dotted curve of \cref{fig:cavity}
corresponds to the maximum deviation between \cref{eq:newexpr} and the
fitted value of \(\Delta\nu\) along our \(1M_\sun\) track (as in the
yellow point of the bottom panel). In principle, the phase function
\(\kappa\), as we have defined it, also becomes discontinuous in the
neighbourhood of \(\nu = \max \nu_\mathrm{ac} \sim \nu_\text{max}\), and
its derivative also becomes singular. A proper accounting should permit
these two singularities to cancel each other out. However, this is
difficult to handle numerically.

It is possible that these difficulties can be resolved by a more refined
analysis. For instance, strictly speaking, the asymptotic behavior of
the WKB wavefunction near such a second-order stationary point should be
given by the parabolic cylinder function \citep{gough_elementary_2007}.
However, such an analysis is fairly involved, and may not as readily
yield a simple (and easily computed) expression like \cref{eq:newexpr}.
Alternatively, one might imagine choosing a set of dynamical variables
such that the expression for the cutoff frequency is always singular
near the center of the model \citep{gough_linear_1993}, so that this
transition point does not emerge. Again, doing so would require a more
detailed analysis than we have performed, which we leave beyond the
scope of this paper.

\hypertarget{rgb-and-red-clump-stars}{%
\subsubsection{RGB and Red Clump stars}\label{rgb-and-red-clump-stars}}

\begin{figure}
\centering
\includegraphics{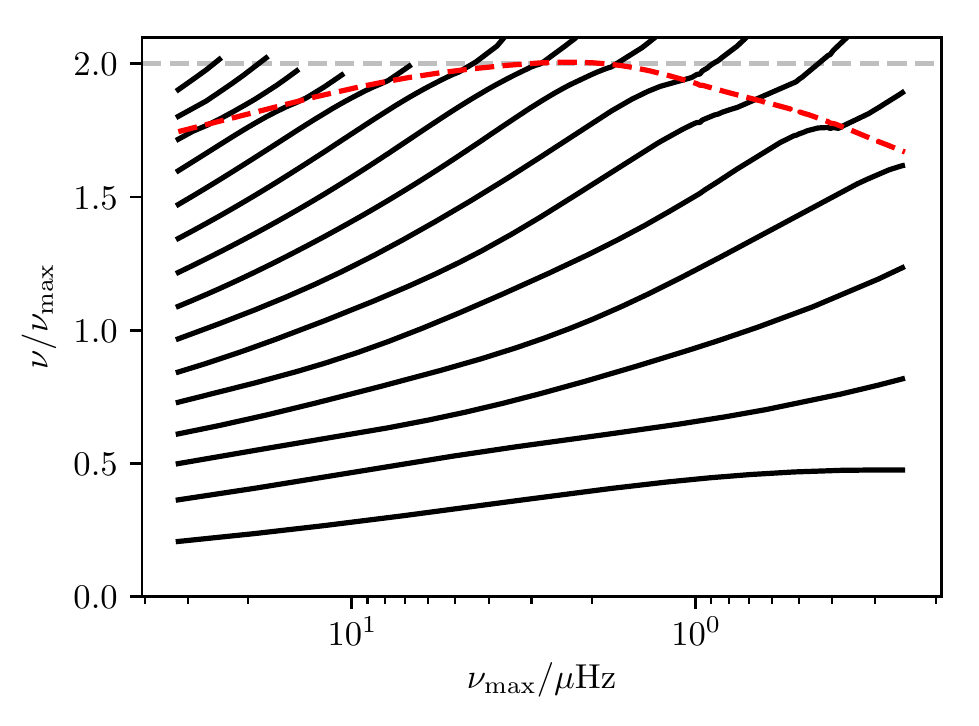}
\caption{Frequencies of radial modes (in units of \(\nu_\text{max}\))
for stellar models on the ascending red giant branch from the
\(1 M_\Sun\) evolutionary track, with modes of the same \(n\) at
different models being joined with solid lines. The maximum acoustic
cutoff frequency in the atmosphere is shown with the red dashed line;
only modes below this dashed line have a confined mode cavity, and so
only these are included in the fit for \(\Delta\nu\). As the age of the
star increases (with decreasing effective temperature), this cutoff
decreases relative to \(\Delta\nu\), and so the number of modes
available to be included in the fit decreases
discontinuously.\label{fig:upperlim}}
\end{figure}

In the lower two panels of \cref{fig:quadrature}, we show the log-ratios
of frequency-separation estimators for ascending RGB (middle panel) and
descending RGB and red clump stars (lower panel). Once again, our
expression in \cref{eq:newexpr} deviates much less from the fitted
\(\Delta\nu\) than do both the scaling relation prediction and the
sound-travel time. However, we note the emergence of terrace-like
features in these deviation plots, which ultimately originate from the
method by which we compute a fitted value for \(\Delta\nu\). In
particular, since we have included all modes with frequencies lower than
the maximum acoustic cutoff frequency in the least-squares fit, and
since the acoustic cutoff frequency changes relative to \(\Delta\nu\)
over the course of stellar evolution, the number of modes used in the
fit changes also changes discontinuously over the course of the track,
as illustrated in \cref{fig:upperlim}. These terraces are therefore a
property of our benchmarking methodology, rather than being a feature of
our asymptotic estimator.

\begin{figure}
\centering
\includegraphics{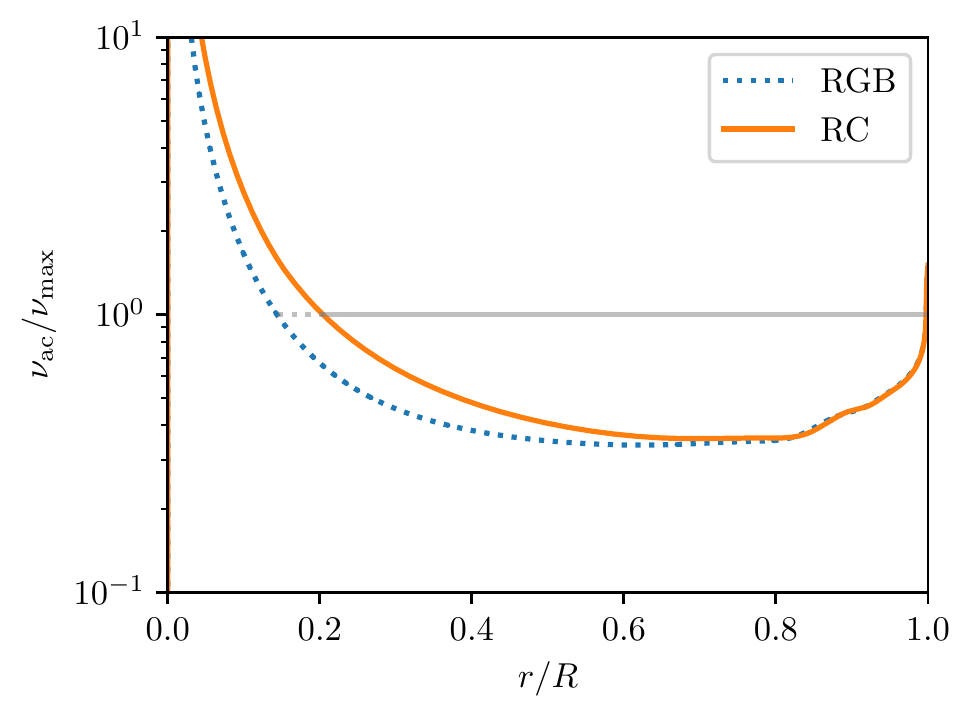}
\caption{Acoustic cutoff frequency (in units of \(\nu_\text{max}\)) for
two \(1 M_\sun\) stellar models on the ascending RGB (dotted line) and
red clump (solid line) with the same \(\log g\) (and therefore the same
radii). While their acoustic structure is very similar near the
envelope, the inner turning points of the modified radial \(p\)-mode
cavity are different, as are the estimated values of
\(\Delta\nu\).\label{fig:cavity2}}
\end{figure}

In addition, we observe that while the scaling relation overpredicts the
large frequency separation for ascending RGB stars, it instead slightly
underpredicts it for red clump stars. Since the use of the scaling
relation is (implicitly) a homology argument
\citep{belkacem_seismic_2013}, this indicates that the acoustic
structure of red clump stars differs significantly from red giant stars.
On this basis, \citet{miglio_asteroseismology_2012} propose the use of
different correction factors for the scaling relation for first-ascent
red giant and red clump stars, precisely to account for these structural
differences.

In terms of our formulation, we show in \cref{fig:cavity2} the radial
dependence of the acoustic cutoff frequency for two stellar models ---
one red giant and one red clump --- with identical masses and radii.
While their acoustic structure is very similar in the outer parts of the
star, we see that their modified radial mode cavities have different
inner turning points. Accordingly, our integral estimator correctly
returns different estimates for \(\Delta\nu\) for these different types
of stars.

\hypertarget{dependence-on-model-atmospheres}{%
\subsubsection{Dependence on Model
Atmospheres}\label{dependence-on-model-atmospheres}}

\begin{figure*}[p]
\centering
\includegraphics{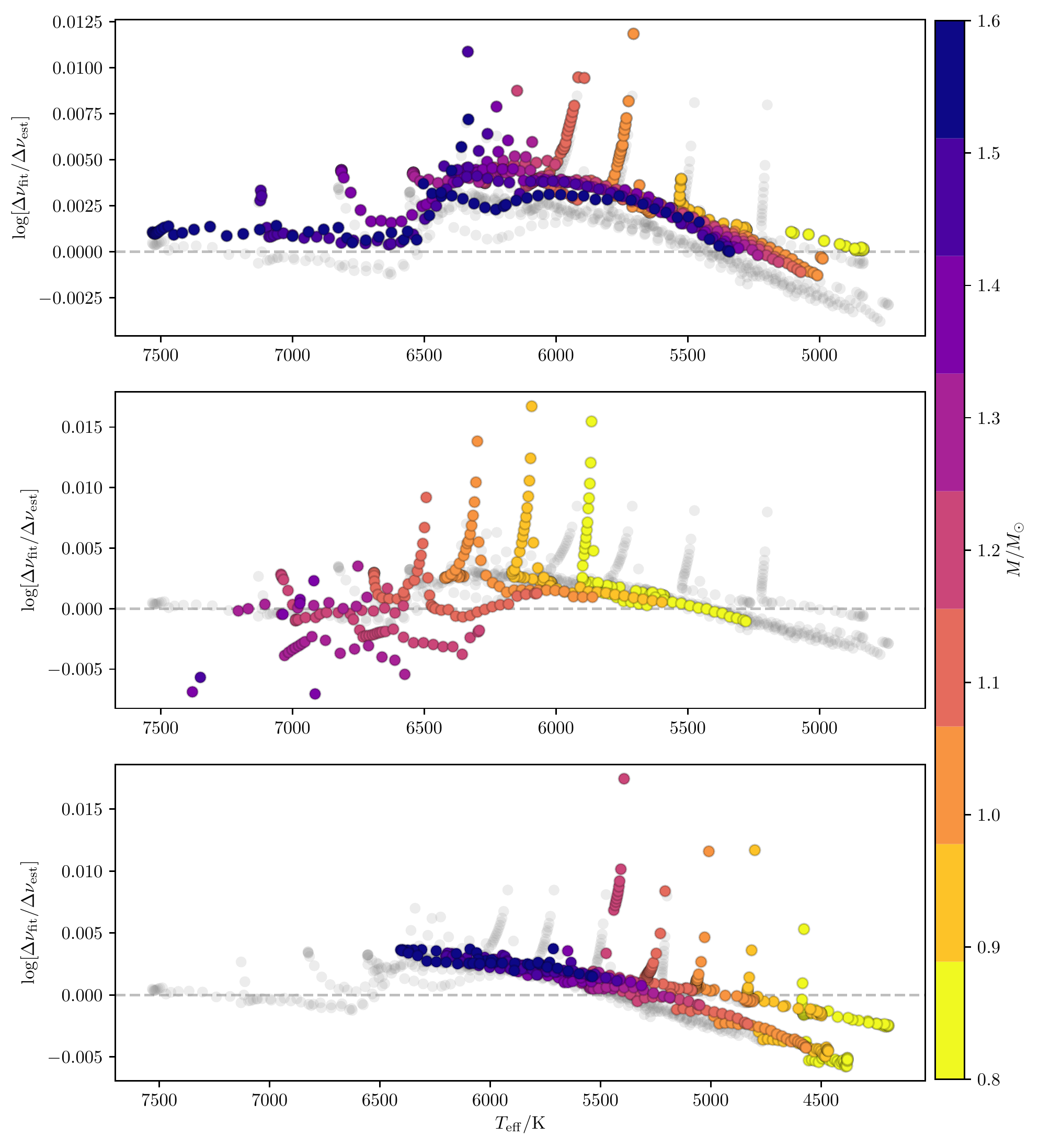}
\caption{Log ratio of fitted \(\Delta\nu\) versus our integral estimator for main-sequence/subgiant models with different physics and/or fundamental parameters. The grey points in the background correspond to the colored points in \cref{fig:quadrature}, at solar metallicity and with Eddington-grey atmospheres. \textbf{Top:} Models with Krishna-Swamy atmospheres instead of Eddington-grey atmospheres; \textbf{Middle:} Low-metallicity models (i.e. $[\mathrm{Fe/H}] = -0.5$), excluding models with no outer convection zone (where we would not expect global oscillations to be excited at all); \textbf{Bottom:} High-metallicity models (i.e. $[\mathrm{Fe/H}] = +0.5$).\label{fig:quadrature2}}
\end{figure*}

The models previously discussed were constructed with Eddington-grey
atmospheres. To examine the dependence of our integral estimator on
atmospheric conditions, we also constructed models using Krishna-Swamy
model atmospheres, with \(Y_0\) and \(\alpha_\text{MLT}\) calibrated
separately. Once again, we show the log ratios between our estimator and
the fitted \(\Delta\nu\) in the top panel of \cref{fig:quadrature2}.
Surprisingly, there does not appear to be any significant difference in
the structure of the residuals. We interpret this to indicate that the
remaining deviations that persist in either case stem from either an
insufficiently high order of approximation, or from issues with our
formulation that become significant only in the interior of the star,
rather than the atmosphere.

One possible such inadequacy in our formulation could be our numerical
treatment of the acoustic cutoff frequency --- for example,
\citet{gough_linear_1993} constructs a different expression for the
acoustic cutoff frequency that takes the sphericity of the star into
account, which only becomes significant deep in the stellar interior.
This is also, therefore, where we expect the expression that we have
used to be the most inaccurate. Since a large contribution to the value
of our integral expression comes from singularities (integrable or
otherwise) at the turning points, it is possible that the remaining
error could potentially be considerably reduced, or eliminated entirely,
by using an expression that is more accurate in the interior.

\hypertarget{dependence-on-metallicity}{%
\subsubsection{Dependence on
metallicity}\label{dependence-on-metallicity}}

To investigate dependences on composition, we also construct stellar
models with metallicities \([\mathrm{Fe/H}]=\pm0.5\), showing the
log-ratios between the fitted frequency separation and our estimator in
the bottom two panels of \cref{fig:quadrature2}. We show only models
possessing outer convection zones. We note that, aside from some
translation along the apparent residual curve, the qualitative features
of these residuals are once again essentially unchanged.

\begin{figure*}[htb]
\centering
\includegraphics{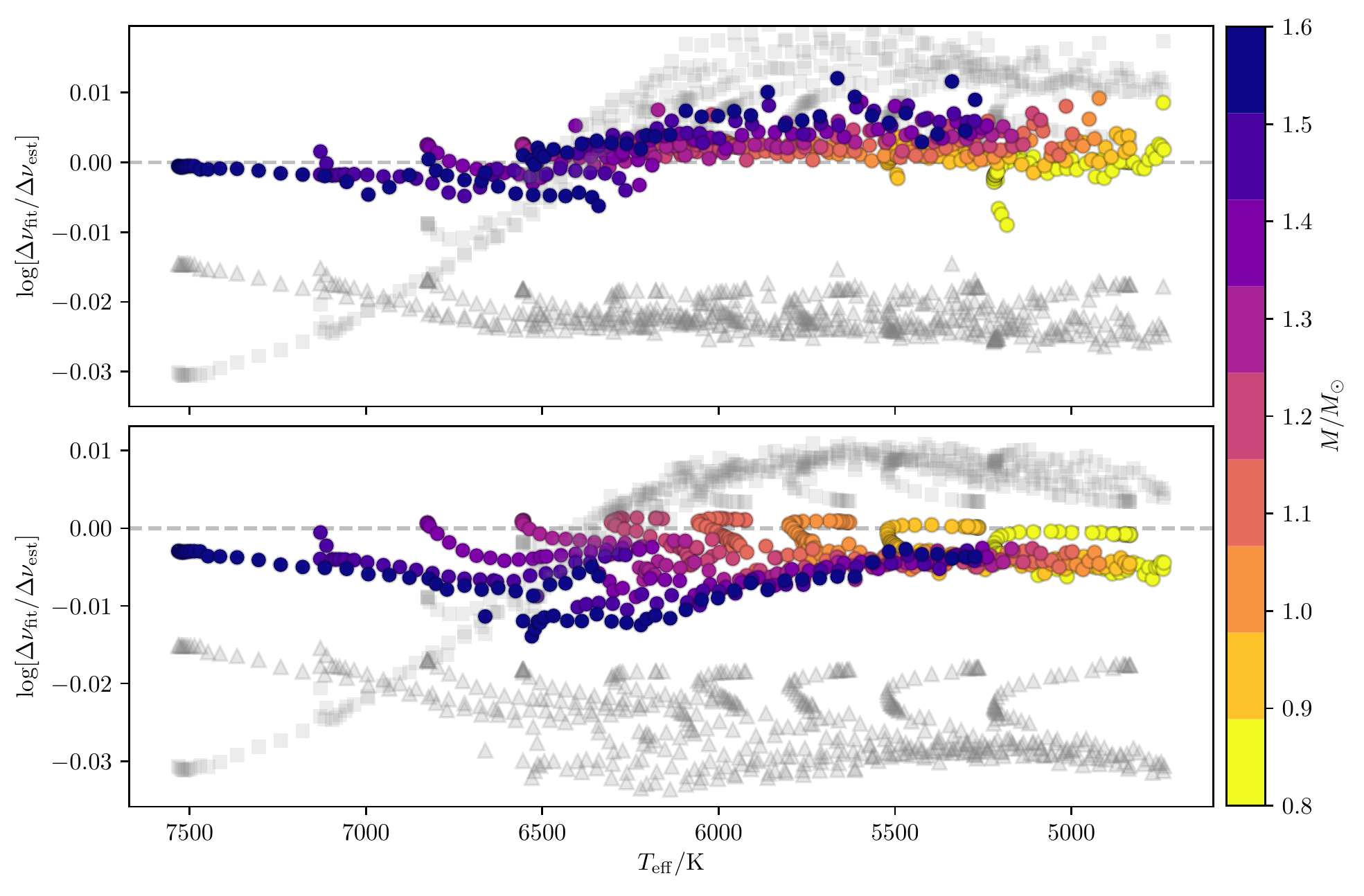}
\caption{Log ratio of fitted \(\Delta\nu\) versus various estimators, with the same coloring and markers as in \cref{fig:quadrature}. \textbf{Top:} Ratios for $\Delta\nu$ fitted from, and integral expressions with, $l=1$. \textbf{Bottom:} The same with $l=2$.\label{fig:quadrature3}}
\end{figure*}

\hypertarget{deltanu-from-nonradial-modes}{%
\subsubsection{\texorpdfstring{\(\Delta\nu\) from nonradial
modes}{\textbackslash{}Delta\textbackslash{}nu from nonradial modes}}\label{deltanu-from-nonradial-modes}}

Finally, if we relax our previous restriction to radial modes, we arrive
at a similar integral expression \begin{equation}
    \Delta\nu_l \sim \left( 2 \int_{r_1}^{r_2}{\mathrm d r \over c_s}{1 - {S_l^2 N^2 \over \omega^4}\over \sqrt{1 - {\omega_\text{ac}^2 \over \omega^2}- {S_l^2 \over \omega^2}\left(1 - {N^2 \over \omega^2}\right)}}\right)^{-1}\label{eq:nonradial}
\end{equation} for the large separation of modes with \(l\ne 0\). To
test this expression, we evaluate \begin{equation}
    \Delta\nu_l \sim \nu_{n+1, l} - \nu_{n, l}
\end{equation} by least-squares fitting of \(\nu_{n,l}\) against \(n\)
for \(l=1\) and \(2\) on the same set of stellar models as used in the
first panel of \cref{fig:quadrature}, again including all modes with
frequencies below the maximum acoustic cutoff frequency. Again, we plot
the log ratio of these against the corresponding integral estimates
(this time from \cref{eq:nonradial}) in \cref{fig:quadrature3}. We see
that for main-sequence stars, our estimator also adequately reproduces
the fitted value of \(\Delta\nu_l\). However, there is a considerable
amount of scatter for stars that have evolved off the main sequence, as
well as a clear systematic deviation for \(l=2\).

We attribute the above to difficulties in constructing a well-defined
average \(\Delta\nu\) for nonradial modes with evolved stars. For
\(l=1\), avoided crossings emerge as the star evolves through the
subgiant phase, and mixed-mode propagation (i.e.~evanescent coupling to
a g-mode cavity) becomes possible
\citep{scuflaire_nonradial_1974, osaki_nonradial_1975, aizenman_avoided_1977, deheuvels_constraints_2010},
with further mode splitting emerging when \(\Delta\nu\) becomes
sufficiently small. This g-mode coupling also changes the frequencies of
observed \(l=2\) mixed modes relative to pure p-modes. Mixed modes are
described by two quantum numbers \(n_p\) (the radial quantum number
\(n\) of the p-mode being coupled) and \(n_g\)
\citep{unno_nonradial_1989}; in computing \(\Delta\nu_l\), we have
selected for each value of \(n_p\) the frequency of the corresponding
mode with the lowest inertia. However, this does not necessarily
correspond to the eigenfrequency of a pure p-mode with that value of
\(n_p\).

\begin{figure}
\centering
\includegraphics{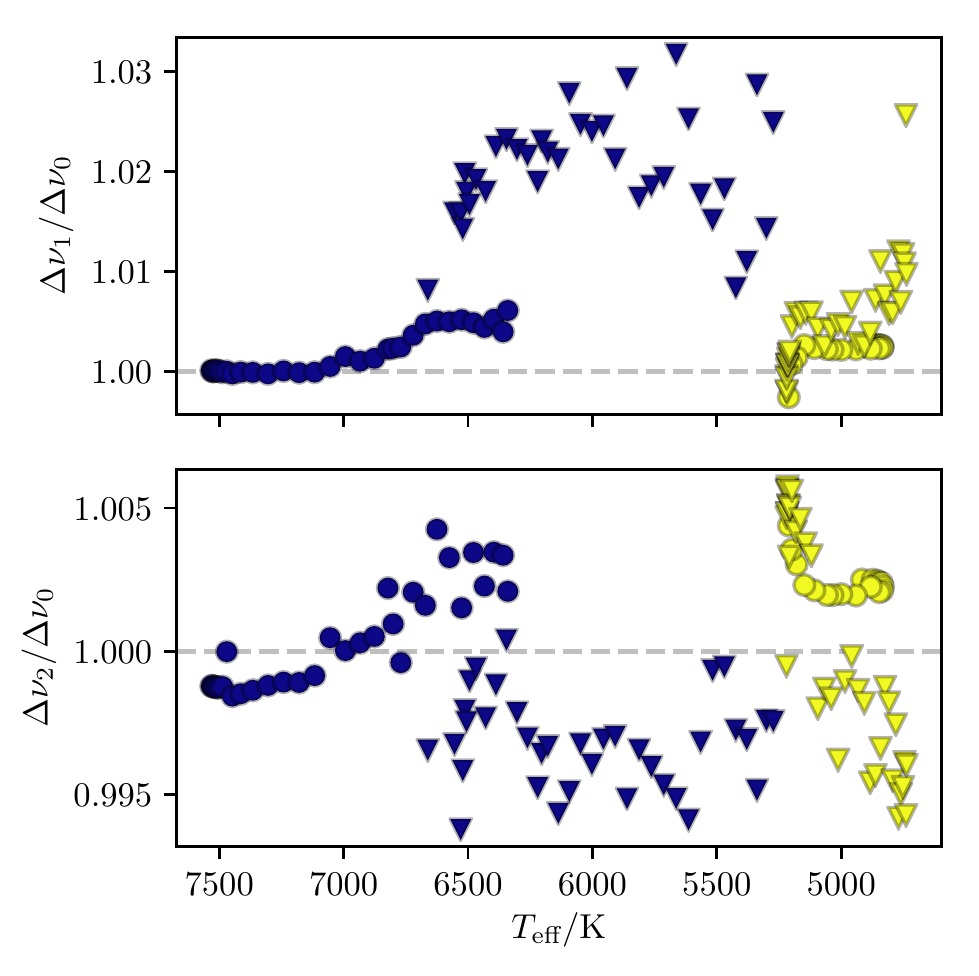}
\caption{Ratios of \(\Delta\nu_l\) to \(\Delta\nu_0\) for \(l=1\) (top)
and \(l=2\) (bottom), for evolutionary tracks of stellar masses \(0.8\)
and \(1.6~M_\sun\), with colors corresponding to the earlier figures
\cref{fig:quadrature}. Main-sequence models are denoted with circles,
while subgiants (with core hydrogen mass fractions \(X<10^{-4}\)) are
shown with inverted triangles.\label{fig:nonradial}}
\end{figure}

We illustrate this problem in \cref{fig:nonradial}, where we plot the
ratio of the least-squares fitted values of \(\Delta\nu_{l}\) to
\(\Delta\nu_{0}\) for \(l=1,2\). It is clear that this ratio is
relatively constant (although not necessarily unity) near the main
sequence, but shows large variations --- both systematic and in terms of
scatter between timesteps --- for evolved stars. Much of the scatter in
\cref{fig:quadrature3} is a result of these rapid variations in
\(\Delta\nu_l\), rather than originating from our modified estimator.
While our formulation cannot explain the remaining systematic
differences for evolved stars (as in the bottom panel of
\cref{fig:quadrature3}, for \(l=2\)), we nonetheless note that these
have qualitatively the same morphology as the relative differences
between the sound-travel time and the fitted value of \(\Delta\nu\).
Again, this likely indicates limitations of our numerical formulation of
the acoustic cutoff frequency; a careful accounting \citep[as
in][]{gough_linear_1993} shows that it, too, depends on the degree \(l\)
(in a somewhat complicated fashion), which is not included in our
numerical integration.

\hypertarget{conclusions}{%
\section{Conclusions}\label{conclusions}}

We present a modified asymptotic expression, \cref{eq:newexpr}, as an
estimator for \(\Delta\nu\), the large frequency separation. We find
that our estimator, evaluated at \(\nu_\text{max}\), describes the large
frequency separation (as obtained by fitting \(\nu\) against \(n\)) more
accurately than both the scaling relation and the classical asymptotic
estimator, which is the sound travel time. While the latter can be
modified to more closely match the fitted value by adjusting the
integration domain somewhat arbitrarily, the turning points of our
integral emerge naturally from the theoretical formulation, and do not
suffer such ambiguity. This result appears to hold good with little
variation with respect to choice of model atmosphere, and modifications
to the model metallicity also do not substantially change the
qualitative features of the residual deviation. The insensitivity of the
residual differences to the choice of model atmosphere indicates that
they originate from theoretical issues pertaining to the interior of the
stellar models, rather than the surface, as is usually assumed
\citep{hekker_tests_2013}.

We also find that our integral expression becomes singular at some point
during main-sequence turnoff; this failure mode is ultimately a
consequence of the failure of the WKB regime under these conditions. We
show that these singular points occur during a transition between two
extreme regimes of asymptotic behavior, owing to structural evolution
yielding a qualitative change in the inner turning point of the WKB
integral. We argue that this provides theoretical justification for
separately calibrated scaling relations for stars at different
evolutionary stages.

Moreover, our naive application of the WKB approach appears insufficient
to completely describe the behavior of \(\Delta\nu\) as fitted from
nonradial modes, particularly away from the main sequence. However, this
may be improved with a more detailed numerical implementation. In any
case, such a fit is observationally ill-defined away from the main
sequence.

These limitations notwithstanding, we propose the use of this integral
expression as an alternative to both the scaling relation and the
sound-travel time for estimating \(\Delta\nu\), although care should be
taken to avoid the singular points where it fails. This expression is
particularly well-suited to such a use when fitting stellar models to
individual mode frequencies, in that the closeness of a model's
\(\Delta\nu\) to observed values can be used as a decision criterion for
whether or not said model is sufficiently optimal as to warrant
expending additional computational resources to calculate its individual
mode frequencies. An accurate asymptotic estimation of \(\Delta\nu\)
will considerably reduce the size of this expensive search space.

\acknowledgements

The authors thank the referee, Dr.~B. Mosser, for the very helpful
comments and suggestions. This work was partially supported by NSF grant
AST-1514676 and NASA grant NNX16AI09G to S.B.

\software{SciPy stack \citep{scipy}, \texttt{MESA} \citep{paxton_modules_2015}, \texttt{GYRE} \citep{townsend_gyre:_2013}.}

\bibliography{biblio.bib}

\end{document}